\documentclass[aps,prb,twocolumn,letterpaper,superscriptaddress,amsmath,amssymb]{revtex4-2}

\usepackage{hyperref}
\usepackage{graphicx}
\usepackage{dcolumn}
\usepackage{bm}
\usepackage{float}
\usepackage[abs]{overpic}
\usepackage[T1]{fontenc}
\usepackage{txfonts}
\usepackage{physics}

\newcommand{\lco}{La$_2$CuO$_4$}
\newcommand{\ybco}{YBa$_2$Cu$_3$O$_6$}
\newcommand{\scoc}{Sr$_2$CuO$_2$Cl$_2$}
\newcommand{\kB}{k_\text{B}}
\newcommand{\Ep}{E_\text{p}}
\newcommand{\Eg}{E_\text{g}}
\newcommand{\Rp}{R_\text{p}}
\newcommand{\deltap}{\delta_\text{p}}

\newcommand{\tpk}{0}
\newcommand{\nex}{n_\text{ex}}
\newcommand{\vint}{v}
\newcommand{\Fsat}{F_\text{s}}
\newcommand{\nsat}{n_\text{s}}
\newcommand{\nh}{n_\text{h}}
\newcommand{\nd}{n_\text{d}}
\newcommand{\Ch}{C_\text{h}}
\newcommand{\taug}{\tau_\text{g}}
\newcommand{\gammah}{\gamma_\text{h}}
\newcommand{\gammad}{\gamma_\text{d}}

\begin{document}

\title{Many-body recombination in insulating cuprates}

\author{Derek G. Sahota}
\affiliation{Department of Physics, Simon Fraser University, V5A~1S6, Canada}
\author{Ruixing Liang}
\affiliation{Canadian Institute for Advanced Research, Toronto, M5G~1Z8, Canada}
\affiliation{Department of Physics and Astronomy, University of British Columbia, V6T~1Z4, Canada}
\author{M. Dion}
\affiliation{Institut Quantique, Regroupement qu\'{e}b\'{e}cois sur les mat\'{e}riaux de pointe, D\'{e}partement de Physique, Universit\'{e} de Sherbrooke, Sherbrooke, Qu\'{e}bec, Canada J1K 2R1}
\author{Patrick Fournier}
\affiliation{Canadian Institute for Advanced Research, Toronto, M5G~1Z8, Canada}
\affiliation{Institut Quantique, Regroupement qu\'{e}b\'{e}cois sur les mat\'{e}riaux de pointe, D\'{e}partement de Physique, Universit\'{e} de Sherbrooke, Sherbrooke, Qu\'{e}bec, Canada J1K 2R1}
\author{Hanna A. D{\k{a}}bkowska}
\affiliation{Brockhouse Institute for Materials Research, McMaster University, Hamilton, L8S~4M1, Canada}
\author{Graeme M. Luke}
\affiliation{Canadian Institute for Advanced Research, Toronto, M5G~1Z8, Canada}
\affiliation{Department of Physics and Astronomy, McMaster University, Hamilton, L8S~4M1, Canada}
\author{J. Steven Dodge}
\affiliation{Department of Physics, Simon Fraser University, V5A~1S6, Canada}
\affiliation{Canadian Institute for Advanced Research, Toronto, M5G~1Z8, Canada}

\date{\today}

\begin{abstract}
We study the pump-probe response of three insulating cuprates and develop a model for its recombination kinetics. The dependence on time, fluence, and both pump and probe photon energies imply many-body recombination on femtosecond timescales, characterized by anomalously large trapping and Auger coefficients. The fluence dependence follows a universal form that includes a characteristic volume scale, which we associate with the holon-doublon excitation efficiency. This volume varies strongly with pump photon energy and peaks near twice the charge-transfer energy, suggesting that the variation is caused by carrier multiplication through impact ionization.
\end{abstract}

\maketitle

\section{Introduction}\label{sec:intro}
Optical excitations and the processes that return them to equilibrium are well understood in conventional semiconductors, where the Coulomb interactions among carriers may be treated perturbatively~\cite{Haug2004,Landsberg1991}. But qualitatively new physics can emerge as the interaction strength increases, such as the well-known Mott gap in the excitation spectrum of correlated insulators~\cite{Basov2011}. Recent research has also shown that the optical recombination processes of interacting systems may exhibit qualitative differences from their more weakly interacting counterparts~\cite{Aoki2014,Basov2011,Giannetti2016}. For example, experiments on cold fermionic atoms in optical lattices show that interactions suppress the recombination rate between empty and doubly-occupied sites, causing it to fall exponentially with increasing on-site repulsion energy~\cite{Strohmaier2010,Sensarma2010,Eckstein2013a}. By contrast, interactions may enhance such recombination in antiferromagnetic insulators, by opening new magnetic channels for decay~\cite{Lenarvcivc2013,Lenarvcivc2014}. Beyond these two-particle recombination processes, interactions may also enhance three-particle processes such as Auger recombination and its inverse process, impact ionization~\cite{Manousakis2010,Coulter2014,Gomi2014,Werner2014,Wang2015,Holleman2016,wais2018}.

Insulating cuprates have served as an important model system for these studies. Their equilibrium properties have been studied extensively because of their relationship to high-temperature superconductors~\cite{Kastner1998}, and numerous measurements have established that their recombination rates exceed those of conventional semiconductors by more than two orders of magnitude~\cite{Landsberg1991,Matsuda1994,Ashida2002,Okamoto2010,Okamoto2011,Petersen2017,Miyamoto2018,Novelli2014}. This extraordinarily rapid recombination is thought to be mediated by magnetic excitations, through a process shown schematically in Fig.~\ref{fig:pescupratefig1}(d,e)~\cite{Lenarvcivc2013,Lenarvcivc2014},  which can dissipate the gap energy more efficiently than phonons could. This process is expected to be relevant in the cuprates even at temperatures well above the N{\'{e}}el temperature, because they exhibit two-dimensional magnetic correlations up to a temperature scale $T\sim J/\kB\sim 1000~\text{K}$,  where $J$ is the magnetic Heisenberg energy~\cite{Kastner1998}.

But there is a problem with the magnetically mediated recombination picture that remains unresolved. The recombination rate between free carriers should show a strong dependence on the excitation density that is not observed, so it was postulated that the recombination occurs via an exciton instead~\cite{Lenarvcivc2013,Lenarvcivc2014}. This assumption conflicts with experiment, since measurements at both terahertz and mid-infrared frequencies show evidence that the recombination involves free carriers~\cite{Okamoto2010,Okamoto2011,Petersen2017}. Here, we develop a kinetic model that resolves this tension, based on measurements of the fluence dependence of the pump-probe response. Following Shockley, Read, and Hall (SRH)~\cite{Shockley1952,Hall1952}, we break the recombination process into two single-particle steps---carrier trapping, followed by recombination---but with much higher rates than found in conventional semiconductors. By fitting this model to our measurements, we identify an additional many-body recombination channel that we associate with an Auger process, which we also find to be anomalously fast. We anticipate that similar kinetics operate in other strongly interacting systems.

An important prediction of our kinetic model is that the leading edge of the pump-probe response should exhibit an apparent shift to earlier times as the fluence increases, which our experiments confirm. This effect is really a form of nonlinear distortion, in which the peak response saturates and grows more slowly with fluence than the onset of the response. Despite its nonlinear origin, however, the shift is \emph{linearly} proportional to fluence at low light levels---demonstrating that nonlinearities can remain relevant even in an experimental regime that appears to be linear.  Earlier experiments have examined the fluence dependence of the pump-probe response in insulating cuprates at similar time scales and probe wavelengths, but did not examine this temporal reshaping~\cite{Okamoto2011, Novelli2014}.

We also demonstrate that fluence dependence offers a way to distinguish the effects of photoexcited charge carriers from those of photoexcited bosonic excitations on the pump-probe response in strongly correlated nonequilibrium systems. Earlier experiments have emphasized temporal and spectral features to accomplish this, for example by associating oscillatory features in the pump-probe signal with an impulsive bosonic response, or by associating photoinduced changes in the terahertz and midinfrared conductivity with fermionic charge carriers~\cite{Okamoto2010,Okamoto2011,Petersen2017,Miyamoto2018}. But experimentally it remains challenging to identify the physical origins of spectral and temporal changes in these systems, so the fluence dependence represents a potentially valuable new way to clarify these.

\section{Experiment}\label{sec:exp}
We studied three insulating cuprates, \ybco\ (YBCO6), \scoc\ (SCOC), and \lco\ (LCO), with two-color optical pump-probe spectroscopy, all at room temperature. We studied single crystals, synthesized through standard methods~\cite{Muller-Buschbaum1977,Liang2006,Dabkowska2010}, of all three materials, as well as a thin film of LCO deposited on a (LaAlO$_3$)$_{0.3}$(Sr$_2$AlTaO$_6$)$_{0.7}$ substrate using pulsed laser deposition. We used two optical parametric amplifiers to generate synchronized 200-fs pump and probe pulses with photon energies $\Ep$ and $E$, respectively, that are independently tunable over 1.65--2.90~eV. We imaged the pump beam through a 1~mm aperture to illuminate a uniform area with a typical 10\%--90\% width of ($0.70 \pm0.05$)~mm at fluence $F$, which was actively controlled by rotating a zero-order achromatic $\lambda/2$ waveplate followed by a linear polarizer. We focused the weaker probe to a $(55 \pm 5)~\muup\text{m}$ (1/$e^2$) spot within the illuminated region to monitor the normalized reflectance change as a function of time, probe photon energy, pump photon energy, and fluence, $\Delta R(t,E;\Ep,F)/R(E)$ (where we use a semicolon to separate the probe parameters $t,E$ from the pump parameters $\Ep, F$).

\section{Results and discussion}\label{sec:resultsdisc}
\subsection{Spectral and temporal response}\label{sec:spectemp}
\begin{figure}
  			\centering   		    
			\includegraphics[width=\columnwidth]{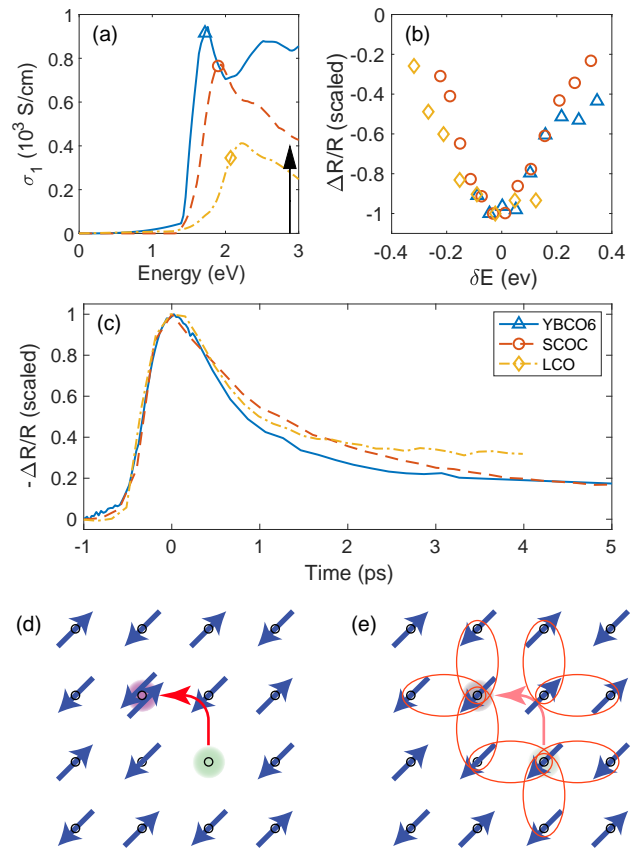}
			\caption{(Color online)~Linear and nonlinear optical response of insulating cuprates at room temperature. (a)~Optical conductivity for YBCO6 (blue solid line and triangle)~\cite{Zibold1993}, SCOC (red dashed line and circle)~\cite{Zibold1996} and LCO (yellow dot-dashed line and diamond)~\cite{Falck1992}. Markers indicate the probe energy $E_0$ with the peak pump-probe response for each material (1.72, 1.90, and 2.07~eV for YBCO6, SCOC, and LCO, respectively), and the arrow indicates the pump photon energy $\Ep=~2.88$~eV used for panels (b) and (c). (b)~Pump-probe response spectra, $\Delta R(\tpk,E;\Ep,F)/R(E)$,  with $\Ep=~2.88$~eV and $F=~(0.83\pm 0.05)~\text{mJ}/\text{cm}^2$, normalized to $|\Delta R(\tpk,E_0;\Ep,F)/R(E_0)|$ (0.096, 0.19, and 0.052 for YBCO6, SCOC, and LCO, respectively) and plotted as a function of $\delta E = E - E_0$ for each material. (c)~Time dependence of the peak response, $\Delta R(t,E_0;\Ep,F)/R(E_0)$, for the pump conditions in (b), normalized to $\Delta R(\tpk,E_0;\Ep,F)/R(E_0)$. (d,e)~Schematic of a recombination process in an antiferromagnetic insulator~\cite{Lenarvcivc2013,Lenarvcivc2014}. Blue arrows represent electrons and their spin direction. When an unoccupied site recombines with  a doubly-occupied site (d), it dissipates the Mott gap energy into the spin system (e).}
			\label{fig:pescupratefig1}
\end{figure}

Figure~\ref{fig:pescupratefig1} shows the basic linear and nonlinear optical response of the materials we studied. All three have a broad peak at 1--2~eV in the linear conductivity shown in Fig.~\ref{fig:pescupratefig1}(a), which marks the charge-transfer gap transition between oxygen and copper orbitals. Photoexcitation above this gap causes the spectrum to broaden and shift to lower energies, which in turn causes the absorption to increase below the gap and decrease above it~\cite{Matsuda1994,Okamoto2010,Okamoto2011,Novelli2014}. This can be seen as a dip in the differential probe reflectance spectra [i.e., $\Delta R(t,E;\Ep,F)/R(E)$ with fixed $t=\tpk$, $\Ep$ and $F$] of Fig.~\ref{fig:pescupratefig1}(b), which are shifted by a material-dependent energy $E_0$ and divided by the peak magnitude $|\Delta R(\tpk,E_0;\Ep,F)/R(E_0)|$ to emphasize the similarity among the three materials. Here and elsewhere, we define $t=\tpk$ by the peak response at fixed $E$ and $E=E_0$ by the peak response (within our probe bandwidth) at fixed $t=\tpk$, both in the low-fluence limit. Figure~\ref{fig:pescupratefig1}(c) shows how the response evolves with time: it decays by about a factor of two during the first picosecond after photoexcitation, then decays more gradually afterwards. These measurements are qualitatively consistent with other studies~\cite{Matsuda1994,Ashida2002,Okamoto2010,Okamoto2011,Novelli2014,Petersen2017}.

\subsection{Saturation with fluence}\label{sec:sat}
\begin{figure}
  			\centering   		    
			\includegraphics[width=8.6cm]{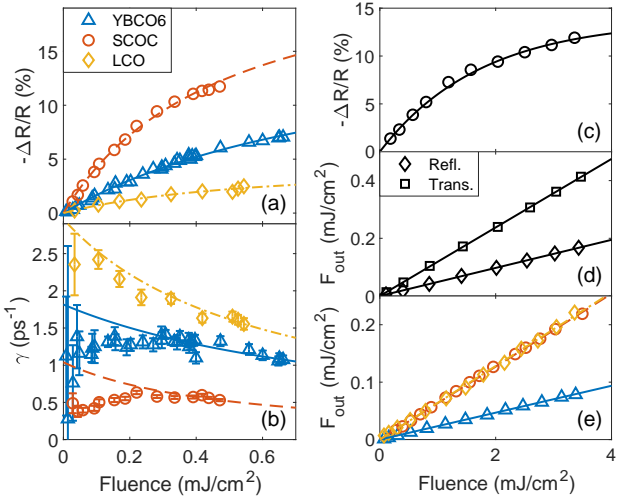}				
			\caption{(Color online) Fluence dependence of (a)~the peak amplitude $\Delta R(\tpk,E_0,\Ep)/R(E_0)$ and (b)~the decay rate $\gamma = -[{\dd(\Delta R)/\dd{t}}]/\Delta R$ evaluated at $t\approx 0.2$~ps for YBCO6 and SCOC with $\Ep =2.88$~eV and single-crystal LCO with $\Ep=2.69$~eV, all at room temperature. Lines in (a) represent least-squares fits to Eq.~(\ref{eq:saturation}) with $\Fsat = (0.96\pm 0.08), (0.49\pm 0.05),$ and $(0.6\pm 0.2)\ \text{mJ/cm}^2$ and $\alpha = 0.18\pm 0.01, 0.25\pm 0.02,$ and $0.05\pm 0.01$ for YBCO6, SCOC, and LCO, respectively. Lines in (b) show $\gamma(F) = \gamma_0 (1 + F/\Fsat)^{-1}$,  the decay rate expected if the saturation shown in (a) were entirely due to nonlinearity in $\nex\to\Delta R$, with fluence-independent kinetics. For each material $\gamma_0$ is chosen to make the curve pass through the measurement at the highest fluence. (c)~The differential reflectance of thin-film LCO saturates with fluence, even as (d)~the transmitted pump fluence ($\square$) and reflected pump fluence ($\diamond$) both remain linear with incident fluence. (e)~Fluence reflected from single-crystal samples of YBCO6, SCOC, and LCO at $E_\text{p} = 3.10~\text{eV}$ as a function of incident pump fluence, with lines showing linear fits. Markers and line styles for (a,b,e) are indicated in the legend of (a).}

			\label{fig:pescupratefig2}
\end{figure}

\begin{figure}
  			\centering   		    
			\includegraphics[width=\columnwidth]{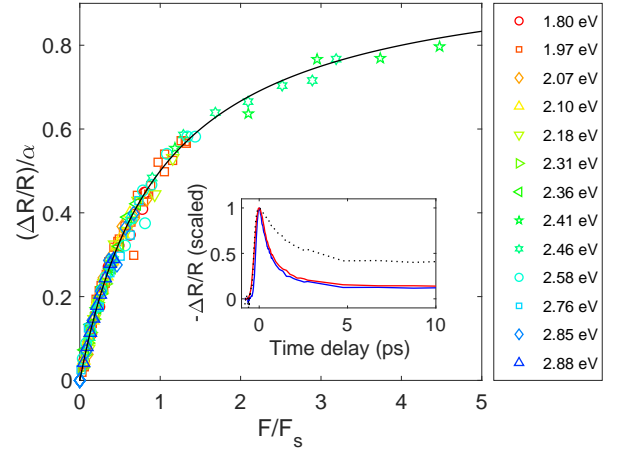}			
			\caption{(Color online) Scaling plot of the peak differential reflectance of YBCO6 at $E=1.70$~eV for various pump photon energies at room temperature. Differential reflectance measurements were fit with Eq.~(\ref{eq:saturation}) to obtain estimates of $\alpha$ and $\Fsat$ for each value of $\Ep$, then normalized by $\alpha$ and plotted as a function of normalized fluence $F/\Fsat$. The black curve shows Eq.~(\ref{eq:saturation}) using the same procedure. The inset shows the differential reflectance with $\Ep = 2.46$~eV at $E=1.70$~eV as a function of time (normalized to the peak value) for $F=0.4\Fsat$ (blue) and $F=2.8\Fsat$ (red), where $\Fsat = 0.38$~mJ/cm$^2$; the black dotted curve shows the normalized response expected for $F = 2.8\Fsat$, given $\Delta R(t)/R$  at $F=0.4\Fsat$ and assuming $\bm{\rho}$ independent of $\nex$.}
			\label{fig:pescupratefig3}
\end{figure}

Figure~\ref{fig:pescupratefig2}(a) shows our central observation: in all three insulating cuprates and at all values of $\Ep$ that we have studied, the peak differential reflectance $\Delta R(\tpk,E_0;\Ep,F)/R(E_0)$ saturates with pump fluence and can be fit well with the empirical model
\begin{equation}
\frac{\Delta R(\tpk,E_0;\Ep,F)}{R(E_0)} = \frac{\alpha F/\Fsat}{1 + F/\Fsat},
\label{eq:saturation}
\end{equation}
where $\alpha$ and $\Fsat$ are fit parameters that vary with $\Ep$. Figure~\ref{fig:pescupratefig3} demonstrates that this functional dependence is universal over a wide range of $\Ep$. We have also found that $\Delta R/R$ has a nonlinear dependence on $F$ at probe energies away from $E=E_0$, but the detailed dependence is then complicated by the fact that the differential probe spectrum is also time-dependent, as we will discuss in a subsequent publication. By focusing here on the peak probe energy $E=E_0$, we minimize the sensitivity of our analysis to time-dependent spectral shifts, since $\partial(\Delta R/R)/\partial E|_{E=E_0} = 0$.

Similar fluence dependence has been observed in other pump-probe measurements on insulating cuprates but remains unexplained~\cite{Novelli2014, Okamoto2011, Petersen2017}. If we express the differential reflectance as a function of the excited-state distribution, $\Delta R\left\{\nex(F)\bm{\rho}[\nex(F),t]\right\}$, where $\bm{\rho}$ is a (suitably normalized) vector of occupation numbers that describes the kinetics, then there are three possible sources for nonlinearity in $F\to\Delta R$: nonlinearity in the absorption process, $F\to \nex$; nonlinear dependence of the reflectance on the initial excitation density $\nex$, $\nex\to\Delta R$; or nonlinearity in the excited-state kinetics, $\nex\to\bm{\rho}$.  We examine each in turn.

Optical saturation can can occur when an absorption process is bleached by Pauli blocking, which effectively excludes some fraction of the material from participating in further absorption. But as Fig.~\ref{fig:pescupratefig2}(c,d) shows, the reflected and transmitted pump fluence of our thin-film  LCO sample remain completely linear at fluences well above those that show saturation in the probe differential reflectance. And as Fig.~\ref{fig:pescupratefig2}(e) shows,  the pump reflectance of our LCO, SCOC, and YBCO6 single-crystal samples remain nearly constant to within our uncertainty at fluences up to 3~mJ/cm$^2$, well above $\Fsat$ for each of them. Any nonlinear pump absorption mechanism, including saturated absorption, two-photon absorption, and excited-state absorption, can not explain the observed saturation in $\Delta R/R$.

Nor can the saturation be explained by nonlinearity in $\nex\to\Delta R$ alone. If this were the case, we would expect the time dependence of $\Delta R/R$ to become distorted as $F$ increases through $\Fsat$, saturating with $F$ near the peak while remaining proportional to $F$ away from it, where $\nex$ is in the linear regime. But as the inset to Fig.~\ref{fig:pescupratefig3} shows, the time dependence in YBCO6 at $F = 2.8\Fsat$ is nearly identical to that at $F=0.4\Fsat$. It also clearly disagrees with the behavior expected if $F\to \nex\to \Delta R$ were the dominant path to nonlinearity, shown as a dotted black curve, which we obtain by assuming linear kinetics $\bm{\rho}(t)$ that are independent of $\nex$, so that $\Delta R(t;F) = \Delta R[\nex(F)\bm{\rho}(t)]$ is separable in $F$ and $t$.

We further test the separability of $\Delta R(t;F)$ in Fig.~\ref{fig:pescupratefig2}(b), which shows the initial decay rate $\gamma = -(\text{d}\Delta R/\text{d}t)/\Delta R$ as a function of $F$ for all three compounds. If $\Delta R(t;F) = \Delta R[\nex(F)\bm{\rho}(t)]$, Eq.~(\ref{eq:saturation}) implies $\gamma(F) = \gamma_0(1 + F/\Fsat)^{-1}$. We show this model dependence for each material in Fig.~\ref{fig:pescupratefig2}(b), taking $\Fsat$ from the fits in Fig.~\ref{fig:pescupratefig2}(a) and choosing $\gamma_0$ so that the model matches the measurement at the highest fluence. The model deviates significantly from the measurements in all three materials. This is most noticeable in YBCO and SCOC at low fluence, where the measured $\gamma(F)$ increases with increasing $F$ while the model $\gamma(F)$ steadily decreases with increasing $F$. Similar differences are observable in LCO, though they are less pronounced.

This leaves nonlinearity in the kinetics, $F\to\bm{\rho}$, as the only other possible source of saturation. Just as in weakly-interacting semiconductors, we expect each optical absorption process in a correlated insulator to create an electron-hole pair that then evolves through additional kinetic processes, including electronic thermalization, recombination, and lower-energy boson production. In principle, both bosonic and fermionic excitations should contribute to $\Delta R/R$ and could saturate with increasing fluence. But any model of the fluence dependence will also have implications for the dependence on the probe spectrum and the dependence on time, so we can use the joint dependence on all three experimental parameters---fluence, probe spectrum, and time---to discriminate among theoretical alternatives. This approach complements earlier work that relied primarily on spectral and temporal signatures to interpret the pump-probe response~\cite{Okamoto2010, Okamoto2011, Novelli2014}.

Each electron-hole pair initially carries excess kinetic energy $\Delta E = \Ep-\Eg$, where $\Eg$ is the gap energy. In the absence of recombination and electron-boson interaction processes, by equipartition the resulting (nondegenerate) electron-hole plasma will thermalize at a temperature $\kB\Delta T = \Delta E/2$ above equilibrium, independent of the excitation density. Consequently, we can immediately exclude purely electronic thermalization as the source of fluence nonlinearity. We can also exclude processes in which hot electrons relax by producing lower-energy phonons, since phonon-phonon interactions are generically weak. And while magnetic excitations are strongly nonlinear, especially for the spin-$1/2$ antiferromagnetism of the cuprates, our experimental results point to fermionic excitations as the dominant source of nonlinearity, with many-body recombination as the mechanism for producing it.

A comparison of $\Delta R/R$ at the charge-transfer gap and the photoconductivity $\Delta\sigma$ at terahertz and mid-infrared frequencies provides strong evidence that the nonlinearity is in the fermionic channel~\cite{Okamoto2011, Petersen2017}. Despite widely separated probe frequencies, the time dependence of $\Delta R/R$ and $\Delta\sigma$ are remarkably similar and their peak response saturates at approximately the same fluence level, indicating a common origin. Since $\Delta\sigma = \Delta(ne^2\tau/m)$ is directly proportional to the charge carrier density $n$ but is only indirectly related to the boson density through the scattering time $\tau$, any saturation mechanism that does \emph{not} involve $n$ would require an unlikely coincidence to produce $\Delta R(t;F)/R\propto\Delta\sigma(t;F)$ over the range in $t$ and $F$ that we observe.

Furthermore, the inset to Fig.~\ref{fig:pescupratefig3} provides independent evidence that fermion kinetics is the dominant influence on $\Delta R(t;F)/R$ in YBCO6 for all $t< 10~\text{ps}$. If the response were dominated by bosonic excitations instead, we would expect them to thermalize at an elevated temperature $\Delta T_\text{th}$ well within this 10~ps time window. The large phonon specific heat at room temperature implies that $\Delta T_\text{th}\lesssim 10~\text{K}$ even at the highest fluences employed here~\cite{Petersen2017}, so we would expect to see the sublinear dependence of $\Delta R(t=\tpk;F)/R$ cross over to a linearly proportional relationship $\Delta R(t\gg\tpk;F)/R\propto \Delta  T_\text{th}\propto F$ as a function of time. Yet $\Delta R/R$ retains the same sublinear dependence on $F$ at $t=10~\text{ps}$ as seen at $t=\tpk$.

\begin{figure}
  			\centering   		    
			\includegraphics[width=8.6cm]{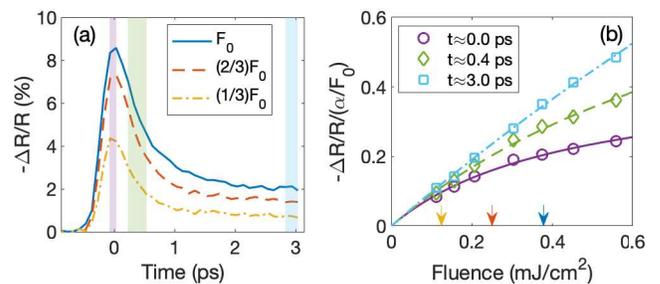}				
			\caption{(Color online) Joint variation of $\Delta R/R$ with fluence and time in LCO with $\Ep = 3.10~\text{eV}$ and $E = E_0 = 2.07~\text{eV}$. (a) Time dependence for equally-spaced values of fluence, with $F_0 = 450~\text{mJ/cm}^2$. (b) Fluence dependence of $\Delta R/R$, averaged over the three temporal ranges shown as colored bands in (a) and normalized so that the response at each fluence has unit slope in the limit $F\rightarrow 0$. Curves show fits to Eq.~(\ref{eq:saturation}) and arrows indicate the fluences shown in (a).}
			\label{fig:pescupratefig4}
\end{figure}

In contrast with YBCO6, in LCO and SCOC we see $\Delta R/R$ cross over with increasing $t$ from a sublinear dependence on $F$ to a linearly proportional dependence on $F$, as shown for LCO in Fig.~\ref{fig:pescupratefig4}. We associate this behavior with a bosonic contribution that becomes more prominent as the fermion contribution decays, which occurs on a time scale of approximately 1~ps in LCO, 10~ps in SCOC, and much greater than 10~ps in YBCO6. In light of this variation, we focus on YBCO6 to develop a kinetic model for the photoexcited charge carriers, since the bosonic contribution to $\Delta R/R$ over our probe bandwidth is weakest in this material.

\subsection{Kinetic model}
The simplest kinetic model that describes our measurements involves first-order SRH decay rates $\gammah$ and $\gammad$ for oxygen holes (holons) and doubly-occupied copper sites (doublons), respectively, together with a third-order Auger decay process with coefficient $\Ch$. Given an energy-dependent holon-doublon excitation efficiency $\eta(\Ep)$ and an excitation rate $g(t) = g_0\Gamma(t)$ with peak amplitude $g_0$ and duration $\taug$, the rate equations for the holon and doublon densities $\nh$ and $\nd$, respectively, are
\begin{align}
\frac{d\nh}{dt} &= \eta(\Ep)g_0\Gamma(t) - \gammah \nh - \Ch \nd \nh^2, \label{eq:rateeqh}\\
\frac{d\nd}{dt} &= \eta(\Ep)g_0\Gamma(t) - \gammad \nd- \Ch \nd \nh^2. \label{eq:rateeqd}
\end{align}
The photocarrier mobility is low so we neglect carrier diffusion~\cite{Petersen2017}, and for simplicity we assume $\eta(\Ep)=1$ until the end of this section, where we explicitly discuss the $\Ep$ dependence. At low $g_0$ the Auger decay is unimportant, and the overall decay rate is limited by the smaller of $\gammah$ and $\gammad$, which we take to be $\gammah$ for definiteness. In this framework, $\gammad$ represents the rate at which defects trap doublons, and $\gammah$ represents the magnetically-mediated recombination rate for a free holon with a trapped doublon. As $g_0$ increases, the Auger decay rate rapidly overtakes the first-order decay processes and causes the photoexcitation density $\nex = \nh + \nd$ to saturate at a characteristic density $\nsat\sim\sqrt{\gammad/\Ch}$. Normally we would expect the Auger decay channel to cause the overall decay rate to grow rapidly with increasing $g_0$, in contradiction with the results in Fig.~\ref{fig:pescupratefig2}(b). But if $\gammad\taug\gg 1$, then the Auger channel is shut off by the rapid decay in $\nd$, allowing the overall decay rate to remain nearly constant at $\gammah$. Indeed, this is precisely what is observed in measurements with shorter pulses~\cite{Miyamoto2018}. 

\begin{figure}
  			\centering   		    
			\includegraphics[width=8.6cm]{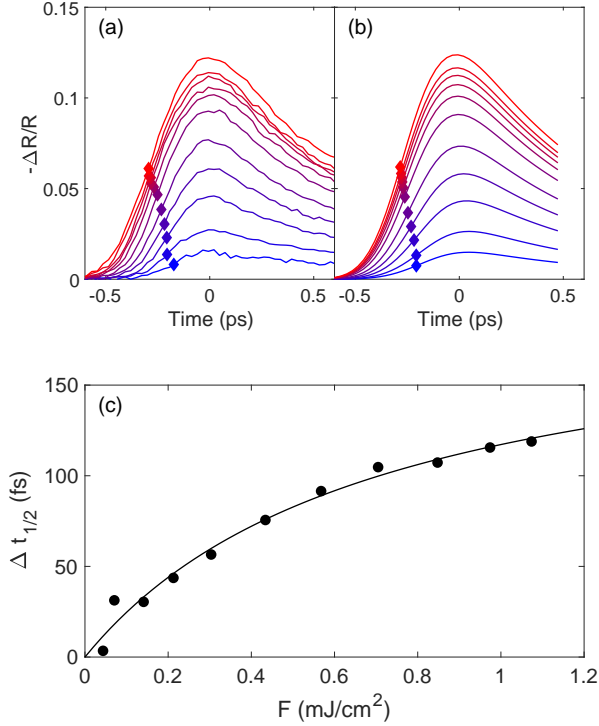}				
			\caption{(Color online) (a)~Differential reflectance at room temperature of YBCO6 at $E=1.70$~eV with $\Ep = 2.46$~eV as $F$ increases from 0.04~mJ/cm$^2$ (blue) to 1.07~mJ/cm$^2$ (red). (b)~Model results with $\eta=1$, $\gammah = 1.3~\text{ps}^{-1}$, $\gammad = 32~\text{ps}^{-1}$, $\Ch = 7.8\times 10^{-26}~\text{cm}^6/\text{s}$, and $\chi = 280e^{-2.75i}~\text{Cu}$. Markers in (a) and (b) indicate the half-maximum point on the leading edge, which shows a temporal shift $\Delta t_{1/2}$ to earlier times. (c)~Fluence dependence of $\Delta t_{1/2}$ for the points shown in (a), fit with $\Delta t_{1/2} = \alpha_{1/2}(F/F_{1/2})(1 +  F/F_{1/2})^{-1}$ (black line).}
			\label{fig:pescupratefig5}
\end{figure}

We can now describe the full dependence of $\Delta R/R$ on $F$, $\Ep$, and $t$ by using Eqs.~(\ref{eq:rateeqh}) and (\ref{eq:rateeqd}) to determine $\nex$, then computing $\Delta R/R$ from the local nonequilibrium permittivity $\epsilon_\text{neq} = \epsilon_\text{eq} + \chi \nex$, where $\chi$ is the susceptibility to the nonequilibrium excitation density. We assume that $g(t)$ decays exponentially with depth over the equilibrium penetration depth; this produces a nonexponential dependence in $\nex$ as it saturates, which we account for when computing $\Delta R/R$~\cite{born1999}. Figure~\ref{fig:pescupratefig5} compares this model to measurements of $\Delta R/R$ for YBCO6 as a function of both time and fluence, with model parameters obtained from a maximum-likelihood fit near the peak, $-0.6 \leq t/\text{ps} \leq 0.5$. We determined the uncertainty in $\Delta R/R$ for these fits directly from at least 10 repetitive measurements at each value of $\Delta t$, using the 90\% trimmed mean standard error to reduce the influence of outliers. We also assumed a 10\% uncertainty in each fluence measurement, applied as an overall scale factor to all values of $\Delta t$ at a given fluence.

This fit yields $\Ch \approx 8\times 10^{-26}~\text{cm}^6/\text{s}$, an Auger coefficient that is four orders of magnitude larger than in GaAs~\cite{Strauss1993}. And while systematic uncertainties in the model and in the laser pulse parameters limit both the overall fit quality and the accuracy of the individual fit parameters, the model clearly captures the qualitative features of the data, including a subtle shift in the leading edge of the response, shown as a function of fluence in Figure~\ref{fig:pescupratefig5}(c). We can associate this shift with nonlinear pulse distortion, in which the leading edge of the response grows linearly with fluence at $t\lesssim \tpk$, then saturates at $t\gtrsim \tpk$ as Auger recombination depletes $\nex$. The quality of this agreement, and the large value of $\Ch$ necessary to achieve it, supports theoretical predictions of enhanced Auger and impact ionization processes in Mott insulators~\cite{Manousakis2010,Coulter2014,Gomi2014,Werner2014,wais2018}.

\begin{figure}
  			\centering   		    
			\includegraphics[width=8.6cm]{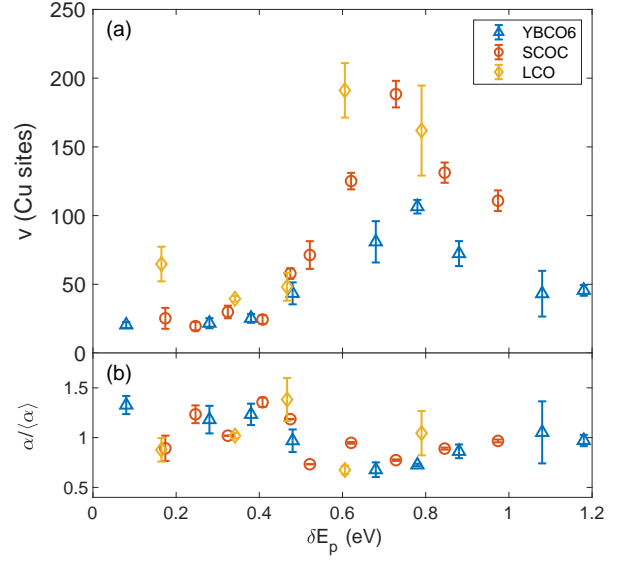}				
			\caption{(Color online) Saturation parameters at room temperature as a function of excitation energy $\delta \Ep = \Ep - E_0$. (a) Effective excitation volume $\vint$. (b) Saturated differential reflectance amplitude $\alpha$, normalized for each material to its average $\langle\alpha\rangle$ over all measured $\Ep$.}
			\label{fig:pescupratefig6}
\end{figure}

We may also use $\Fsat$ to determine a characteristic interaction volume, $\vint = \Ep\deltap/[(1-\Rp)\Fsat]$, where $\deltap$ and $\Rp$ are the penetration depth and the reflectance, respectively, at the pump energy $\Ep$. Figure~\ref{fig:pescupratefig6} shows how $\vint$ and $\alpha$ vary with $\Ep$. In all three materials $\vint$ has a pronounced maximum just above the charge-transfer energy, while the saturated differential reflectance $\alpha$ is relatively constant over our photoexcitation bandwidth. Thus, while $\vint$ sets the scale for saturation at high fluence, it also determines the limiting behavior at low fluence, $\Delta R(F)/R = \alpha F/\Fsat\propto \alpha\vint  F$.

This behavior motivates the efficiency factor $\eta(\Ep) = g_0/F$ that we have included in Eqs.~\ref{eq:rateeqh} and \ref{eq:rateeqd}. The values of $\vint$ and $\Fsat$ are determined primarily by the nonlinear terms of Eqs.~\ref{eq:rateeqh} and \ref{eq:rateeqd}. These play no role in the behavior at low fluence, so with constant $\eta(\Ep)$ the model incorrectly predicts that $\lim_{F\rightarrow 0}\dd{(\Delta R/R)}/\dd{F}$ is independent of $\vint$ and  $\Fsat$. But when $\eta(\Ep)$ is allowed to vary, it effectively rescales the $F$ dependence while leaving the $\Delta R/R$ scale unchanged, automatically reproducing the observed $\lim_{F\rightarrow 0}\dd{(\Delta R/R)}/\dd{F} \propto\Fsat^{-1}\propto \vint$.

Within this framework, the peak in $\vint(\Ep)$ shown in Fig.~\ref{fig:pescupratefig6}(a) implies a peak in the photocarrier generation efficiency $\eta(\Ep)$ at $\Ep\approx E_0 + 1~\text{eV}\approx 2\Eg$ in all three materials [see Fig.~\ref{fig:pescupratefig1}(a)]. This energy scale strongly suggests a role for impact ionization, the inverse process to Auger recombination, in which a high-energy charge carrier relaxes to the gap energy by exciting an additional electron-hole pair across the gap. This can only occur if the carrier has a kinetic energy of at least $\Eg$, so we expect it to become important when $\Ep\gtrsim 2\Eg$, where we observe the peak in $\vint(\Ep)$. This process will compete with electron-phonon and electron-magnon relaxation, but is expected to dominate in systems with sufficiently strong electron-electron interactions~\cite{Manousakis2010,Coulter2014,Gomi2014,Werner2014,wais2018}.

Equations~\ref{eq:rateeqh} and \ref{eq:rateeqd} thus provide a good qualitative description of all of our experimental results, including the saturation of $\Delta R/R$ with fluence, the weak fluence dependence of the decay rate, the fluence-dependent temporal shift of the $\Delta R/R$ onset, and the dependence of the saturation parameters $\alpha$ and $\vint$ with pump excitation energy $\Ep$. But since the model originates from a rigid-band, independent-electron description of solids that clearly does not apply to the cuprates, the underlying parameters require some reinterpretation. Most importantly, the holon and doublon densities $\nh$ and $\nd$ refer to states that are in principle density-dependent themselves. The fact that they provide a good description of our experiments suggests that this apparent lack of self-consistency is unimportant in practice, presumably because the interactions renormalize the trapping and recombination parameters without changing the underlying kinetic description. An interesting avenue for future work is to explore how this description breaks down, especially as the density of interaction-induced in-gap states increases.

Our interpretation of the trapping parameters $\gammah$ and $\gammad$ is also somewhat different from the original SRH framework. Traps in weakly-interacting semiconductors are associated with bound states of a one-electron potential that remain largely inert with respect to occupation changes in other states, so the trap density may be assumed fixed. These states will become fully occupied at sufficiently high excitation densities, forcing newly generated electron-hole pairs to occupy free-electron states that will then modify the kinetics. We see no evidence for such trap saturation, even though we employ percent-level excitation densities in samples with the highest quality available. But if we instead understand trapping in the cuprates as a process of polaron formation and localization, then the impurity potential necessary to trap a carrier will decrease as the polaron forms, pushing the excitation density for trap saturation above the levels normally found in conventional semiconductors.

Finally, we note that the Auger recombination parameter $\Ch$ is thought to be enhanced by the large on-site Hubbard interaction in strongly interacting systems, in contrast with the long-range Coulomb interactions that produce Auger recombination in conventional semiconductors~\cite{Manousakis2010,Coulter2014,Gomi2014,Werner2014,wais2018,manousakis2019}. In principle, interactions may also enhance $\Ch$ by broadening the spectral function and relaxing the phase-space restrictions imposed by energy and momentum conservation on the scattering amplitude. But this broadening also reduces the spectral amplitude, effectively cancelling any enhancement that could be gained from this effect~\cite{manousakis2019}.

\section{Conclusion}\label{sec:conc}
In summary, we have shown that the insulating cuprates satisfy standard SRH recombination kinetics, but with much higher trapping and recombination rates than found in conventional semiconductors. The associated kinetic equations successfully describe the nonequilibrium response as a function of both time and fluence over a wide range of pump and probe wavelengths, and harmonizes the current theory of magnetically-mediated recombination more effectively with existing experiments. Our results indicate that interactions influence multiple kinetic processes, and might provide an avenue for controlling nonequilibrium behavior in applications.

Our emphasis on the fluence dependence of $\Delta R/R$ complements and clarifies earlier work that focused on its dependence on time and probe energy~\cite{Matsuda1994,Okamoto2010,Okamoto2011,Novelli2014}. As our results demonstrate, the fluence dependence is sensitive to the photon absorption rate, so it can reveal processes that occur on timescales that are much shorter than the duration of the either the pump or the probe. Furthermore, we have shown that the fluence dependence can provide essential guidance for developing a model of the photocarrier kinetics, which can then shape the interpretation of the temporal and spectral response. For example, earlier work typically neglected the possibility that the temporal response could be reshaped by the recombination kinetics as we describe, so it would be useful to revisit them with greater attention to this effect~\cite{Okamoto2011,Novelli2014}.

We can identify several other directions for further development of this work. For example, by examining the probe spectrum as a function of fluence and time we may further disentangle the contributions to $\Delta R/R$ from charge carriers and neutral bosons, as we will discuss in later work. Also, the magnetically-mediated recombination rate should vary exponentially with both the gap energy $\Eg$ and the magnetic interaction energy $J$, but we observe a relatively constant single-particle decay rate across materials that have similar $J$ and a gap energy that varies by 20\%. Similar studies of materials with a wider range of $\Eg/J$ and at variable temperature could help clarify whether this mechanism is sufficient to explain the recombination in the cuprates, and whether it is relevant in other antiferromagnetic insulators. Such studies could also help determine the factors that contribute to the trapping, Auger recombination, and impact ionization rates of Mott insulators, which are currently not well characterized experimentally.

\begin{acknowledgments}
JSD thanks Malcolm Kennett, George Sawatzky, Andrew Millis, Bill Atkinson, and Efstratious Manousakis for helpful discussions. PF, GL, and JSD acknowledge support from NSERC and CIFAR; JSD from the Canada Foundation for Innovation, \href{https://www.westgrid.ca/}{WestGrid}, and \href{www.computecanada.ca}{Compute Canada}; PF from the Canada First Research Excellence Fund; and DGS from an NSERC Alexander Graham Bell Canada Graduate Scholarship.
\end{acknowledgments}

\end{document}